\documentstyle[12pt]{article}

\newcommand{\be}{\begin{equation}}
\newcommand{\ee}{\end{equation}}
\newcommand{\bea}{\begin{eqnarray}}
\newcommand{\eea}{\end{eqnarray}}
\newcommand{\al}{\alpha}
\newcommand{\G}{\Gamma}

\begin{document}

\title{\bf
Next-to-next-to-leading order relation 
between $R(e^+e^-\rightarrow b\bar b)$
and $\Gamma_{\rm sl}(b\rightarrow cl\nu_l)$ 
and precise determination of $|V_{cb}|$
}
\author{
  {\bf A.A.Penin and A.A.Pivovarov}\\
  {\small {\em Institute for Nuclear Research of the
  Russian Academy of Sciences,}}\\
  {\small {\em 60th October Anniversary
  Pr., 7a, Moscow 117312, Russia}}
        }

\date{}

\maketitle

\begin{abstract}
We present the next-to-next-to-leading order relation
between the moments of  the $\Upsilon$ system spectral
density  and the inclusive $B$-meson semileptonic width.
The perturbative series 
for the width as an explicit function of the 
moments is well convergent 
in three consequent orders in
the strong coupling constant
that provides solid and accurate theoretical estimate.
As a result, the uncertainty
of the value of $|V_{cb}|$ Cabibbo-Kobayashi-Maskawa matrix element
is reduced.   

\end{abstract}

\thispagestyle{empty}

\newpage
\noindent 
The inclusive $B$-meson semileptonic width $\Gamma_{\rm 
sl}(b\rightarrow cl\nu_l)$
is  rather a clean place to obtain the value of 
the Cabibbo-Kobayashi-Maskawa (CKM)
matrix element $|V_{cb}|$ (see refs.\cite{BSU,N,U} as a review).
The main uncertainty of this estimate is related to the 
strong dependence of the result on the  $b$-quark pole mass $m_b$
which has to be known with extremely high accuracy to get
a  precise estimate of $|V_{cb}|$. On the other hand
the high moment sum rules for the system of $\Upsilon$
resonances are very sensitive to  $m_b$
and can be used for the precise determination of this 
parameter after the proper treatment of 
Coulomb effects \cite{NSVZ,Vol}. 
In the present paper we construct the direct relation
between the moments of the $\Upsilon$ system spectral
density  and the inclusive $B$-meson semileptonic width up to the  
next-to-next-to-leading (NNL) order of perturbative expansion 
in $\al_s$. In this way we avoid 
the strong dependence of $\Gamma_{\rm sl}$
on   $m_b$ and essentially reduce
the theoretical uncertainty in  $|V_{cb}|$. It is known
that if 
the moments for the 
$\Upsilon$
system and the inclusive semileptonic width
are expressed through the pole mass $m_b$
the perturbative expansions in $\al_s$ for 
both quantities
seem to be divergent.
The heurictic criterium of convergence of perturbative series,
i.e. the requirement that next correction is much smaller than the
previous one, is violated.
However, the explicit dependence of 
the inclusive semileptonic width on 
the pole mass $m_b$ can be removed by 
reexpressing it through 
the moments.
As a result one arrives at the perturbative expansion 
for the quantity $V_{cb}$ which 
converges well up to NNL order in $\al_s$. 
   
The theoretical expression for the inclusive
semileptonic width of $B$-meson
reads
\be
\Gamma_{\rm sl}={G_F^2m_b^5\over 192\pi^3}|V_{cb}|^2F\left(
{m_c^2\over m_b^2}\right)C_\Gamma(\alpha_s)
\label{Gamma}
\ee 
where $F(x)=1-8x-12x^2\ln{x}+8x^3-x^4$.  
The perturbative coefficient $C_\Gamma(\alpha_s)$ is known up to the second
order in $\al_s$ \cite{CM}
\[
C_\Gamma=1-1.67{\al_s(\mu)\over\pi}
-(8.4\pm 0.4)\left({\al_s\over\pi}\right)^2
\]
where $\al_s(\mu)$ is defined 
in $\overline{\rm MS}$ scheme and the normalization point
$\mu=\sqrt{m_bm_c}$ is used. The nonperturbative corrections
to  eq.~(\ref{Gamma}) decrease the width by approximately $5\%$ \cite{U,np}.

 From formula~(\ref{Gamma}) one sees 
that the semileptonic width depends
rather strongly on $m_b$, the b-quark mass, which is the pole mass.
This means that if $m_b$ is taken from some other experiment it should
be determined with great accuracy for reasonable determination of
$|V_{cb}|$. 
Being analysed independently, 
the perturbative series in $\al_s$ for 
physical observables expressed in terms of the pole mass $m_b$
seem not to enjoy a fast explicit
convergence that can lead to large uncertainty of the numerical 
value of $m_b$ extracted from different 
experiments.
However $m_b$ is not an observable and has no
immediate physical meaning. Therefore it can be safely removed from 
relations between physical observables.  
Here we use this strategy for direct
determination of
the mixing angle $|V_{cb}|$ 
reducing theoretical uncertainties to a great extent.
Our analysis consists in 
direct relating the factor $m_b^5$ in eq.~(\ref{Gamma}) to
the fifth moment of the $\Upsilon$ sum rules 
\be
m_b^5
=\left({{\cal M}^{th}_5\over \tilde{{\cal M}}_5^{exp}}\right)^{1\over 2}.
\label{mb}
\ee 
The moments ${\cal M}^{th}_n$ are defined as  normalized derivatives
of  the  $b$-quark vector current polarization function  $\Pi(s)$
\[
{\cal M}^{th}_n= 
\left.{12\pi^2\over n!}(4m_b^2)^n{d^n\over ds^n}\Pi(s)\right|_{s=0}=
(4m_b^2)^n\int_0^\infty{R(s)ds\over s^{n+1}}
\]
where 
\be 
\left(q_\mu q_\nu-g_{\mu \nu}q^2\right)\Pi(q^2)=
i\int dxe^{iqx}\langle 0|Tj_{\mu}(x)j_{\nu}(0)|0\rangle , 
\qquad j_\mu=\bar b\gamma_\mu b
\label{momth}
\ee
and $R(s)=12\pi {\rm Im}\Pi(s+i\epsilon)$. 
The (dimensionful) experimental moments $\tilde{{\cal M}}_n^{exp}$ are generated by 
the function $R_b(s)$ which is the 
normalized cross section
$R_b(s) = \sigma(e^+e^-\rightarrow {\rm hadrons}_{\,b\bar b})/
\sigma(e^+e^-\rightarrow \mu^+\mu^-)$
\be
\tilde{{\cal M}}_n^{exp} = 
{4^{n}\over Q_b^2}\int_0^\infty{R_b(s)ds\over s^{n+1}}.
\label{expmom}
\ee
Here $Q_b=-1/3$ is the $b$-quark electric charge.

The theoretical moment ${\cal M}^{th}_5$ is a dimensionless quantity 
which depends on $m_b$  only 
logarithmically as $\ln{(m_b/\mu)}$
(in a finite order in $\al_s$). 
Thus, the substitution of relation~(\ref{mb}) to eq.~(\ref{Gamma})
substantially reduces its dependence on $m_b$ though introduces 
explicit uncertainty due to ${\cal M}^{exp}_5$.
At the  same time
the  theoretical moment is rather sensitive to
the value of  $\al_s$.

Let us discuss the choice of the moment in 
expression~(\ref{mb}) for the $b$-quark  pole mass
in more detail. 
It is well known that for the high moments 
the ordinary perturbative  expansion in the strong coupling constant
is not applicable and the Coulomb resummation is required.
Recently 
the expansion of the theoretical
moments around the exact Coulomb solution has been obtained 
in the NNL approximation
\cite{KPP,PP,H,MY}.
In our analysis we use the analytical expression of the 
polarization function near threshold obtained in refs. \cite{KPP,PP}.    
The explicit formulae are too large to be presented here.

The experimental moment entering eq.~(\ref{mb}) 
is given by the integral of the spectral density 
eq.~(\ref{expmom}) and is mostly saturated with the contribution of
the first six $\Upsilon$ resonances that leads to the formula
\be
\tilde{{\cal M}}_n^{exp}={4^{n}\over Q_b^2}
\left({9\pi\over \al_{QED}^2(m_b)}
\sum_{k=1}^6{\Gamma_{k}\over M_{k}^{2n+1}}
+\int_{s_0}^\infty\!{\rm d}s {R_b(s)\over s^{n+1}}\right).
\label{expspectr}
\ee
The leptonic widths
$\Gamma_{k}$ and masses $M_{k}$ $(k=1\ldots 6)$ of the resonances
are known with good accuracy \cite{PDG}, 
the electromagnetic coupling constant is renormalized to the 
energy of order $m_b$ with the result 
$\al_{QED}^2(m_b)=1.07 \al^2$ \cite{PDG}.
The rest of the spectrum beyond the resonance
region  for energies larger than
$s_0\approx (11.2~{\rm GeV})^2$ (continuum contribution)
is approximated by the theoretical spectral density
multiplied by the parameter $0.5<t<1.5$ which accounts
for the uncertainty in the experimental data in this energy region.
Note that we assume rather large uncertainty of the continuum 
to be on the safe side however its contribution is essentially
suppressed in comparison with the resonance one 
and the resulting error of the whole quantity in eq.~(\ref{expspectr})
is of the same order
as the uncertainties introduced by the resonance contribution.

Note that $m_b^5$ (or $m_b$) can be formally extracted from an arbitrary
moment of the spectral density. 
The region of allowed $n$ however is quite restricted.   
Indeed, the low moments  cannot
be used in  sum rules because the experimental spectrum is well known
experimentally only for energies close
to threshold due to existence of sharp resonances while the
contribution of the continuum to these low moments is large in
comparison with the resonance contribution. On the other hand
for $n>10$ the  perturbative expansion of the moments around the 
Coulomb solution is strongly divergent. This can be considered 
as a signal that for large $n$ the Coulomb solution is not the 
best zero order approximation \cite{PPp}.  
So $n=5$ seems to be the natural and optimal 
choice. In any case the result
of calculation is almost insensitive to the local variation of $n$
around this value.

  From eqs.~(\ref{Gamma},\ref{mb}) we find for the  mixing angle 
$|V_{cb}|$ 
\be
|V_{cb}| 
=\left({192 \pi^3\over G_F^2}
\G_{\rm sl}\sqrt{M^{exp}_5}{K(\alpha_s,m_b)\over  
F({m_c^2}/{m_b^2})}\right)^{1\over 2}
\label{vcb}
\ee
where the functions
\be
K(\alpha_s,m_b)
={1\over C_\G(\alpha_s)
\sqrt{M^{th}_5(\alpha_s,m_b)}}
\label{Kcoef}
\ee
and $F({m_c^2}/{m_b^2})$ accumulate
theoretical information depending on
$m_b$, $m_c$ and $\alpha_s$.

Function $F(m_c^2/m_b^2)$  introduces
rather large theoretical uncertainty  if masses of $b$- and
$c$-quarks are considered as independent variables.
However there is almost model independent constraint of the form
\be
(m_b-m_c)(1+O(1/m_{b,c}))=\bar m_B - \bar m_D=3.34~{\rm GeV}
\label{const}
\ee
where $\bar m_B=5.31~{\rm GeV}$, $\bar m_D=1.97~{\rm GeV}$ denote
the spin-average meson masses, e.g. $\bar m_B=\frac{1}{4}(m_B+
3m_{B^*})$. We take the value of the nonperturbative corrections to
eq.~(\ref{const}) given in ref.~\cite{BBBG} and use 
$m_b-m_c=3.47~{\rm GeV}$ as the central value 
for numerical estimates. 
With this constraint the function $F(m_c^2/m_b^2)$ 
becomes a function of a single variable
$\tilde{F}(m_b)$. 
Note that in such a setting the $m_b$ dependence of 
the function $\tilde F(m_b)$ 
partly cancels the large $m^5_b$ dependence of the width.
Furthermore using the relation ~(\ref{mb})
to express the $b$-quark pole mass in the argument of the function 
$\tilde F(m_b)$ in terms of the  moments of the spectral 
density results in only logarithmic  
dependence of the right hand side of eq.~(\ref{vcb}) on $m_b$.

Now one can analyze the theoretical factor 
$K(\alpha_s,m_b)/\tilde F\left({m_b}\right)$
numerically order by order in $\al_s$. The result reads
\be
\left({K(\alpha_s,m_b)\over \tilde F(m_b)}\right)^{1\over 2}
=1.234(1+0.102+0.014) .
\label{thseries}
\ee
In the numerical analysis of extracting $|V_{cb}|$
we use the ``world average''
value of the strong coupling constant $\al_s(M_Z)=0.118$
\cite{PDG} and 
the normalization point $\mu\sim m_b$ in the 
expression for the theoretical moment\footnote{
For the numerical estimates it is important to fix 
the allowed range for the normalization point
which is present in the explicit formula of the polarization
function. The naive estimate of the ``physical scale''
of the problem $\mu\sim m_b\al_s$ \cite{Vol,H} 
is not acceptable since the direct calculation of the NNL corrections
shows   that the perturbation theory for the moments blows 
up there \cite{KPP,PP}. The relative weight of the 
NNL order corrections is stabilized at $\mu\sim m_b$ \cite{PPp} 
so in our opinion there is no reason to use the lower 
normalization scales.}. The value of $b$-quark pole
mass in  the fixed order in $\al_s$ with Coulomb resummation
is found from   
eq.~(\ref{mb}). For comparison, the perturbative series
for $m_b$ that follows from eq.~(\ref{mb}) and the series 
for $C_\Gamma$ are
\[
m_b=4.75(1-0.014+0.022) ,
\]
\[
C_\Gamma=1-0.146-0.064 .
\]
Thus we find that in the  
expansions of the theoretical moment (as well as $m_b$ itself) 
and width
expressed in terms of $m_b$ the NNL corrections
are of the order of the next-to-leading ones\footnote{
There is a hypothesis that this fact is a consequence of the 
asymptotic character of the series which leads to the intrinsic
ambiguity in the heavy quark pole mass \cite{ren}.} while    
the perturbative series for the mixing angle,
or the theoretical coefficient eq.~(\ref{thseries}),
converges much better. 

As for numerics, we use the following central values 
for our experimental inputs (see \cite{PDG} for more detail):
\[
{\rm BR}(B\rightarrow X_cl\nu_l)=10.5\%, \quad \tau_B = 1.55~{\rm ps},
\]
\[
\sqrt{M^{exp}_5}=4.51\times 10^{-4}~{\rm GeV}^{-5}.
\]

With these numbers we obtain
the value of the matrix element $|V_{cb}|$
\be
|V_{cb}|=0.0423\left({{\rm BR}(B\rightarrow X_cl\nu_l)
\over 0.105}\right)^{1\over 2}
\left({1.55 {\rm ps}\over \tau_B}\right)^{1\over 2}\times
\label{final}
\ee
\[
\left(1 +0.02{\sqrt{M^{exp}_5}-4.51\times 10^{-4}~{\rm GeV}^{-5}\over 
0.20\times 10^{-4}~{\rm GeV}^{-5}}\right)
\left(1 -0.01{\al_s(M_Z)-0.118 \over 0.006}\right)
\left(1\pm\Delta_{npt}\right)
\]
where the nonperturbative corrections 
are included according to ref.~\cite{U} and $\Delta_{npt}\sim 0.02$
is the uncertainty in the nonperturbative contribution
coming mainly from the uncertainty in the HQET relation between $m_b$
and $m_c$ eq.~(\ref{const}).
The typical scale of uncertainty of key parameters is also
indicated.
Another important source 
of the uncertainty is 
the scale dependence of the theoretical moment 
and the experimental errors 
in the value of $\al_s(M_Z)$ because of rather high
sensitivity of the theoretical moment to $\al_s$.
In fact these uncertainties are closely related since
the scale dependence of ${\cal M}_5^{th}(\al_s(\mu),\mu)$ 
is mainly due to the scale   dependence of $\al_s(\mu)$ 
while the explicit dependence  on $\mu$ is rather
weak. The pointed error bars roughly correspond to the interval 
$\mu = m_b \pm 1~{\rm GeV}$ at 
fixed $\al_s(M_Z)=0.118$.
Note that the result  is almost insensitive to the 
specific number of the moment used for the estimate. 
For example, the  value of  $|V_{cb}|$ matrix element 
changes approximately by $0.2\%$ when the moment
number changes in the interval $5<n<15$. 
  
Now the  advantage of our approach becomes clear --
large errors due to uncertainty in $m_b$ 
is now partly shifted to more direct experimental data.
Furthermore the  terms in the perturbative expansion~(\ref{thseries}) 
decrease rapidly which is a solid indication 
that the  higher  orders corrections 
to the obtained result are small enough.   

The main part of the experimental uncertainty is 
related to the uncertainty in the experimentally measured 
inclusive semileptonic width.
The uncertainty in $M^{exp}_5$ comes mainly from the 
continuum contribution to the moments above $s_0$
and the uncertainties in $\Gamma_{k}$.
Experimental situation is rather dynamic and data are improving fast
that means that the experimental  uncertainties will be smaller 
(see e.g \cite{prosp}).

Our result for the central value of the parameter $|V_{cb}|$ of the CKM mixing
matrix eq.~(\ref{final}) is in a  good   agreement 
with the previous estimate $|V_{cb}|=0.0419$ \cite{U}. 
This value, however,
is somewhat larger than the estimate $|V_{cb}|=0.039\pm 0.002$ 
of ref.~\cite{N}.  
There is no much hope to reduce the uncertainty in the 
nonperturbative contribution. Thus the model 
independent result presented in the paper
provides one with the most reliable and accurate estimate of 
the CKM matrix element $|V_{cb}|$ from
the inclusive $B$-meson semileptonic width.

\vspace{3mm}
\noindent
{\large \bf Acknowledgements}\\[2mm]
We thank J.G. K\"orner and J.H.K{\"u}hn
for discussions
and kind hospitality at Karlsruhe and Mainz Universities
were this work was started. 
A.A.Penin gratefully acknowledges discussions
with K.Melnikov and thanks him for sending results
of ref. \cite{MY} before publication.
This work is partially supported 
by Volkswagen Foundation under contract
No.~I/73611. A.A.Pivo\-varov is  
supported in part by
the Russian Fund for Basic Research under contracts Nos.~96-01-01860
and 97-02-17065. 
The work of A.A.Penin is supported in part by
the Russian Fund for Basic Research under contract
97-02-17065.


\begin{thebibliography}{99}
\bibitem{BSU}  I.Bigi, M.Shifman and N.Uraltsev, Ann.Rev.Nucl.Part.Sci. 
               {\bf 47}(1997)591.

\bibitem{N}    M.Neubert, {\it ``B Decays And the Heavy Quark Expansion''},\\
to appear in the Second Edition of: {\it Heavy
Flavours}, edited by A.J. Buras and M. Lindner (World Scientific, Singapore),
hep-ph/9702375.

\bibitem{U}  N.Uraltsev, {\it ``Heavy Quark Expansion in Beauty 
and its Decays ''}, Lectures given at the International School
of Physics Enrico Fermi {\it ``Heavy Flavour Physics: a Probe of Nature's
Grand Design ''}, hep-ph/ 9804275.

\bibitem{NSVZ} V.A.Novikov {\it et al.}, Phys.Rev.Lett. {\bf 38}(1977)626;\\
               V.A.Novikov {\it et al.}, Phys.Rep. {\bf C41}(1978)1;\\
               M.B.Voloshin, Yad.Fiz. {\bf 36}(1982)247;\\
               M.B.Voloshin and Yu.M.Zaitsev, Usp.Fiz.Nauk
               {\bf 152}(1987)361.

\bibitem{Vol}  M.Voloshin, Int.J.Mod.Phys. {\bf A10}(1995)2865.

\bibitem{CM}   A.Czarnecki, Phys.Rev.Lett. {\bf 76}(1996)4124;\\
               A.Czarnecki and K.Melnikov,  Nucl.Phys. {\bf B505}(1997)65; 
               Phys.Rev.Lett. {\bf 78}(1997)3630; {\bf TTP-98-14}, 
               hep-ph/9804215.

\bibitem{np}   I.Bigi, N.Uraltsev and A.Vainstein, Phys.Lett.  
               {\bf B293}(1992)430;\\
               B.Blok and M.Shifman, Nucl.Phys.   {\bf B399}(1993)441, 459;\\  
               I.Bigi, M.Shifman, N.Uraltsev and A.Vainstein, Phys.Rev.Lett.  
               {\bf 71}(1993)496.

\bibitem{KPP}  J.H.K{\"u}hn, A.A.Penin and A.A.Pivovarov, Preprint 
               {\bf TTP-98-01}, \\ hep-ph/9801356.

\bibitem{PP}   A.A.Penin and A.A.Pivovarov, Preprint 
               {\bf TTP-98-13},  hep-ph/9803363.

\bibitem{H}    A.H.Hoang,  Preprint {\bf UCSD/PTH 98-02}, hep-ph/9803454.

\bibitem{MY}   K.Melnikov  and A.Yelkhovsky, Preprint 
               {\bf TTP-98-17}, hep-ph/9805270. 

\bibitem{PDG}  Particle Data Groop, Phys.Rev. {\bf D54}(1996)1.

\bibitem{PPp}  A.A.Penin and A.A.Pivovarov, in preparation.

\bibitem{BBBG} E.Bagan, P.Ball, V.Braun and G.Gosdzinsky,  Phys.Lett.  
               {\bf B342}(1995)362. 

\bibitem{prosp} S.Stone,  {\it ``Prospects For b Physics in the Next Decade''},\\
                {\bf HEPSY-96-01}, hep-ph/9610305 

\bibitem{ren}  I.Bigi, M.Shifman, N.Uraltsev and A.Vainstein, Phys.Rev.  
               {\bf D50}(1994)2234;\\
               M.Beneke and V.Braun, Nucl.Phys. {\bf B426}(1994)301;\\
               M.Jezabek, M.Peter and Y.Sumino, {\bf HD-THEP-98-10},
               hep-ph/9803337;\\ 
               A.H.Hoang {\it et al.},  Preprint {\bf UCSD/PTH 98-13}, 
               hep-ph/9804227;\\
               M.Beneke,  {\bf CERN-TH/98-120},  hep-ph/9804241.




\end{thebibliography}
\end{document}